\documentclass[11pt]{article}

\usepackage[preprint]{acl}

\usepackage{times}
\usepackage{latexsym}
\usepackage[T1]{fontenc}
\usepackage[utf8]{inputenc}
\usepackage{microtype}
\usepackage{inconsolata}
\usepackage{graphicx, tabularx}
\usepackage{booktabs, multirow} % for borders and merged ranges
\usepackage{soul}% for underlines
\usepackage{xcolor,colortbl} % for cell colors
\usepackage{changepage,threeparttable} % for wide tables
\usepackage{dblfloatfix}
\usepackage{amsmath}
\usepackage{amssymb}
\usepackage[table]{xcolor}
\usepackage{nicematrix}

\usepackage{arydshln}
\usepackage{tcolorbox}
\tcbuselibrary{listings}

\usepackage{color}

\lstdefinelanguage{json}{
  basicstyle=\ttfamily\footnotesize,
  numbers=none,
  stepnumber=1,
  showstringspaces=false,
  breaklines=true,
  frame=none,
  stringstyle=\color{black},
  commentstyle=\color{gray}\itshape,
  morestring=[b]",
  morecomment=[l]{//},
  morecomment=[s]{/*}{*/},
}

\title{
PlanRAG-Audio: Planning and Retrieval Augmented Generation\\
for Long-form Audio Understanding
}
\author{
  \textbf{Masao Someki\textsuperscript{1}},
  \textbf{Chien-yu Huang\textsuperscript{1}},
  \textbf{Siddhant Arora\textsuperscript{1}},
  \textbf{Samuele Cornell\textsuperscript{1}},
  \textbf{Markus M\"uller\textsuperscript{2}},\\
  \textbf{Nathan Susanj\textsuperscript{2}},
  \textbf{Rupak V Swaminathan\textsuperscript{2}},
  \textbf{Grant P Strimel\textsuperscript{2}},
  \textbf{Jing Liu\textsuperscript{2}},
  \textbf{Shinji Watanabe\textsuperscript{1}}\\
\\
  \textsuperscript{1}Language Technologies Institute, Carnegie Mellon University,
  \textsuperscript{2}Amazon AGI
}

\begin{document}
\maketitle
% \begin{abstract}
% Long-form spoken interactions pose unique challenges for large language models (LLMs) due to the massive number of tokens and the need to reason across multimodal cues such as speech, emotion, and acoustic events.
% To address these challenges, we propose \textit{PlanRAG-Audio}, a planning-based retrieval-augmented generation (RAG) framework for efficient long-form audio understanding.
% Unlike conventional RAG systems that perform unstructured retrieval, PlanRAG-Audio first plans which modalities—semantic, paralinguistic, or acoustic—are relevant to a given query, and then executes structured retrieval from a multimodal database.
% This database integrates ASR transcripts, speaker identities, emotion labels, and acoustic events, supporting both symbolic and embedding-based retrieval.
% \textcolor{gray}{Experimental results demonstrate that PlanRAG-Audio substantially improves performance on spoken question answering and event detection tasks, raising BLEU scores from 6.32 to 19.80 and event detection accuracy from 32\% to 45\%, while reducing input length by up to $19.33\times$.}
% PlanRAG-Audio enables scalable application of RAG to long-context audio understanding.
% \end{abstract}
\begin{abstract}
Long-form audio understanding poses significant challenges for large audio language models (LALMs) due to the extreme length of audio sequences and the need to reason over heterogeneous acoustic cues distributed over time, such as speech content, speaker identity, emotion, and sound events.
To address these challenges, we propose \textbf{PlanRAG-Audio}, a planning-based retrieval-augmented generation framework for scalable long-form audio understanding.
Rather than having audio LALMs process entire recordings directly, PlanRAG-Audio explicitly plans which modalities and temporal spans are required for a given query, and retrieves only query-relevant information from a structured text and audio database.
This retrieval planning enables effective reasoning over complex, cross-domain audio queries while substantially reducing the input length passed to the large language models.
Experiments across a wide range of speech/audio retrieval demonstrate that PlanRAG-Audio improves reasoning accuracy and stabilizes performance as audio duration increases by decoupling inference cost from raw audio length.
\end{abstract}

\section{Introduction}

\begin{figure*}[t]
  \centering
  \includegraphics[width=\linewidth]{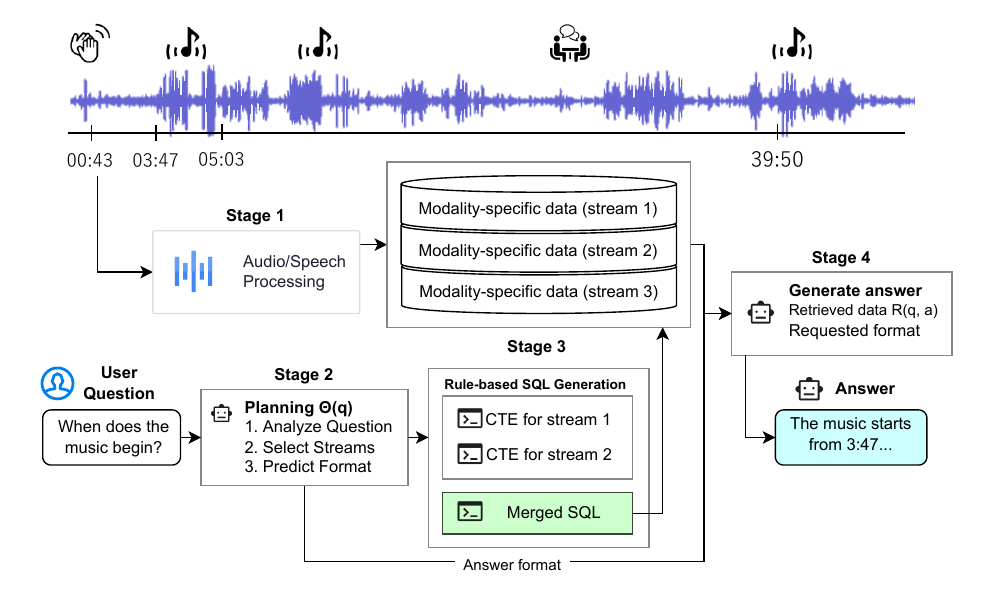}
  \caption{
  Overview of question-driven multimodal retrieval over long-form audio.
Given a user question grounded in audio (derived from audio and speech processing; see Figure~\ref{fig:detailed_audio_ingestion} and Section~\ref{ssec:audio_database_construction}), the system plans the required reasoning steps by analyzing the question, selecting relevant streams, and predicting the requested output format.
For each selected stream, the planned retrieval is compiled into stream-specific Common Table Expressions (CTEs), which are then composed into a unified SQL query via a hybrid LLM–rule-based SQL generator.
The retrieved segments are finally aggregated and passed to the generation model to produce the answer.
  }
  \label{fig:big_picture}
\end{figure*}

Spoken interaction has become a key modality for human--machine communication,
driving the development of large audio language models (LALMs) that jointly reason over linguistic and non-verbal acoustic cues~\cite{audioflamingo,moshi,salmonn,opuslm},
yet long-form speech remains extremely challenging due to the resulting growth in both token length and multimodal complexity.
For example, a one-hour lecture corresponds to roughly 12\,k text tokens but over 100\,k speech tokens based on Gemini~\cite{gemini},
leading to severe computational and memory bottlenecks when conversations span minutes or hours.
Enabling LALMs to efficiently understand and reason over such extended long-form speech data remains an open challenge.

Retrieval-augmented generation (RAG)~\cite{rag} has proven effective in text domains for mitigating hallucination and improving reasoning by retrieving external evidence.
Given the finite context window of large language models (LLM), it is infeasible to include all potentially relevant documents or web data directly within the model input.
RAG addresses this limitation by \emph{selectively retrieving} only the information necessary to answer a user query, allowing the model to focus on a compact and relevant subset of knowledge rather than processing the entire corpus.
As such, RAG serves as a mechanism for efficient information extraction under constrained context length.
Extending this paradigm to audio provides a natural route toward scalable speech, text, and acoustic understanding: instead of examining every audio token in long speech recordings, a system can retrieve only the most relevant semantic, paralinguistic, or acoustic segments needed for reasoning.
Recent work on RAG has shown that explicitly planning the retrieval process before execution can improve efficiency and attribution for complex queries~\cite{planrag,plan_star_rag}.

However, most prior work on long-form audio understanding still relies on automatic speech recognition (ASR) or audio captioning pipelines, which convert speech to text before applying conventional NLP models~\cite{cooral_qa,meetingqa,blab,streamRAG}.
This transcript/text-centric design ignores prosody, speaker variation, and non-verbal acoustic events that are essential for human communication.
Other approaches use audio-text contrastive embeddings~\cite{wavrag,efficient_sqa,Cacophony} or direct audio retrieval~\cite{NAAQA,AttentionAQA}, but these methods remain limited to short clips and fail to account for long-range dependencies across modalities.

Importantly, the difficulty of long-form audio understanding does not arise solely from the increased input length.
As recordings extend over longer time spans, they inherently induce queries that require \emph{compositional reasoning across heterogeneous acoustic cues}.
Such queries often depend on the interaction between spoken content, speaker, and non-verbal sound events distributed over time.
For example, a user may ask to summarize segments of a radio broadcast by jointly considering spoken reports and background acoustic events, while associating each summary with its corresponding time span.
These problems cannot be adequately addressed by text-only representations, such as those obtained via ASR, as their semantics emerge from cross-domain dependencies and long-range temporal structure.
Consequently, effective long-form audio understanding requires a framework that can selectively reason over multiple modalities while preserving temporal structure.

To overcome these limitations, we propose \textbf{PlanRAG-Audio}, a \emph{planning-based retrieval-augmented generation framework} that formulates long-form audio understanding as a structured information retrieval problem.
Given a query, the system first plans which modalities(e.g., spoken content, speaker information, emotional cues, and non-verbal acoustic events), temporal spans, and constraints are required, transforming the task into targeted retrieval over a structured audio database.
It then retrieves only the corresponding information and performs reasoning over the retrieved evidence.
This decomposition avoids redundant processing and enables scalable reasoning over hours of audio without requiring large input contexts.
As illustrated in Figure~\ref{fig:big_picture}, PlanRAG-Audio uses an LLM to produce a retrieval plan, compiles it into structured database queries, and aggregates the retrieved evidence to generate the final answer.

% To overcome these limitations, we propose \textbf{PlanRAG-Audio}, a \emph{planning-based retrieval-augmented generation framework} for long-form audio understanding.
% Instead of retrieving all available information, PlanRAG-Audio first analyzes a user query to plan which modalities (e.g., spoken content, speaker information, emotional cues, and non-verbal acoustic events), which temporal spans, and which retrieval constraints are required to answer the query.
% It then retrieves only the corresponding information from a database.
% This explicit planning stage prevents redundant retrieval, allowing the model to reason efficiently over hours of audio without requiring an extremely large input window.
% As illustrated in Figure~\ref{fig:big_picture}, PlanRAG-Audio uses an LLM to analyze a query and produce a retrieval plan, which is then compiled into structured database queries to retrieve relevant audio metadata, before a generation LLM aggregates the retrieved evidence to produce the final answer.

Our contributions are as follows.
\begin{itemize}
    \item We propose \textbf{PlanRAG-Audio}, a planning-based retrieval-augmented framework for long-form audio understanding that performs \emph{planning before retrieval}.
    \item We show that PlanRAG-Audio enables effective reasoning over long-form audio by retrieving information from multiple modalities while preserving temporal structure.
    \item We demonstrate that PlanRAG-Audio handles a wide range of long-form audio understanding tasks, from base tasks such as QA and diarization to advanced and compositional reasoning tasks, in a zero-shot manner without task-specific prompt engineering or manually crafted SQL queries.
\end{itemize}

\section{Related Work}

% Research related to long-form audio understanding spans multiple research communities, including audio information retrieval, retrieval-augmented generation,
% and benchmark design for extended audio inputs.
% In this section, we review prior work along these three axes and clarify
% how our approach relates to and differs from existing systems.

We review related work on audio information retrieval, retrieval-augmented generation, and long-form audio evaluation.

\subsection{Audio Information Retrieval}

Prior work on audio information retrieval can be broadly categorized into several directions.
One line of research focuses on learning transferable audio representations for tagging and retrieval~\cite{kong2020panns,CLAP2022,Cacophony}.
Another line addresses spoken question answering and spoken document retrieval by combining speech recognition, as well as ASR-free formulations for spoken QA~\cite{lin2021speechdpr,DUAL,NAAQA,AttentionAQA}.
More recently, retrieval-augmented generation has been extended to speech and audio,
integrating audio representations or transcripts with RAG-style pipelines~\cite{min2023speechrag,wavrag,aura,semnani2023wikichat}.
Despite these advances, existing approaches are typically task-specific or limited to short audio segments, and rely on transcript-centric retrieval.
As a result, they lack support for structured, modality-aware retrieval over long-form audio.

\subsection{Retrieval Augmented Generation}

Recent work on retrieval-augmented generation has moved beyond the simple
\textit{retrieve--then--generate} paradigm toward more structured approaches
that incorporate iteration and planning to handle complex queries~\cite{self_rag,ra_isf,planrag,plan_star_rag}.
Related ideas have also been explored in long-context domains such as video understanding,
where hierarchical decomposition and step-by-step retrieval are used to scale RAG to longer inputs~\cite{cadencerag}.
Extending these ideas to long-form audio remains an open challenge,
as effective retrieval must account for speaker changes, emotional dynamics,
and non-speech sound events over long recording.

% Recent advances in RAG have gone beyond the simple \textit{retrieve–then–generate} paradigm,
% exploring iterative and planning-based extensions to handle complex queries in a more structured way.
% %
% Early approaches such as Self-RAG~\cite{self_rag} and RA-ISF~\cite{ra_isf} introduced mechanisms of self-reflection and iterative feedback to improve retrieval accuracy and answer consistency.
% Building on this, PlanRAG~\cite{planrag} and Plan$\ast$RAG~\cite{plan_star_rag} proposed a \textit{plan-then-retrieve} strategy, decomposing queries into sub-queries or directed acyclic graph (DAG) structures to achieve higher efficiency and stronger attribution.
% This trajectory has also extended into video understanding: methods like CadenceRAG~\cite{cadencerag} decompose long-video queries hierarchically and perform step-by-step retrieval and reasoning.
% In this way, RAG frameworks with iterative planning and self-improvement have expanded their scope from text to video.
% Extending these ideas to long-form audio is a natural next step, 
% where the design of RAG systems that incorporate speaker changes, emotional dynamics, and acoustic events represents an inevitable direction for future research.

\subsection{Long-Form Audio Evaluation}
\label{ssec:audio_evaluate}

Research on spoken and long-form audio understanding has evolved from synthetic datasets to realistic, open-domain benchmarks.
Early efforts on acoustic QA (AQA) such as CLEAR~\cite{abdelnour2018clear}, inspired by CLEVR~\cite{johnson2017clevr}, and its extension NAAQA~\cite{NAAQA}, introduced controlled synthetic acoustic scenes to study compositional reasoning over sound attributes.
Later datasets such as Clotho-AQA~\cite{lipping2022clothoaqa} and Audiopedia~\cite{penamakuri2025audiopedia} extended AQA to crowdsourced and knowledge-intensive settings.
However, most of these benchmarks focus on short clips and do not support multi-hour reasoning or multimodal alignment.

More recently, BLAB~\cite{blab} introduced a benchmark for long-form audio understanding, but remains limited in reproducibility and evaluation scope.
In contrast, our work evaluates long-form audio understanding using fully open datasets
and aligns the evaluation domains with tasks commonly studied in recent Interspeech conferences, as detailed in Appendix~\ref{appdx:interspeech_survey}.

\section{Proposed Framework}

In this section, we introduce PlanRAG-Audio by formulating long-form audio understanding as a retrieval planning problem, and describe both the planning mechanism and the underlying audio database design that enable efficient and scalable retrieval over extended audio.

\subsection{PlanRAG-Audio}

\subsubsection{Stage 1: Audio and Speech Processing}
As illustrated in Figure~\ref{fig:big_picture}, PlanRAG-Audio first converts raw audio into a set of modality-specific representations through audio and speech processing.
Each representation corresponds to an independent retrieval stream, such as speech transcripts, speaker segments, sound events, or emotional cues.
These streams are stored in a structured audio database and serve as the basic units
for downstream retrieval.
Details of the audio ingestion process are described
in Section~\ref{ssec:audio_database_construction}.

\subsubsection{Stage 2: Retrieval Planning}
\label{sssec:stage2}
Given a user question $q$, PlanRAG-Audio formulates retrieval as a planning problem, where a planning LLM analyzes the query and produces a structured retrieval plan
$\Theta(q)$.
Because $\Theta(q)$ is generated via constrained decoding under a fixed schema,
the planning stage is effectively deterministic and does not produce
invalid retrieval plans.
This planning step explicitly determines what information should be retrieved before executing any retrieval operations.
Concretely, the retrieval plan $\Theta(q)$ specifies:
1) which modality-specific representations are required;
2) what filters are applied to each stream;
3) how multiple streams are joined;
4) which fields are returned from the merged SQL results;
5) what schema the final generation LLM must follow.
Example~1 shows a simplified example of a retrieval plan $\Theta(q)$.
For clarity, the example omits implementation-specific details and retains only the fields necessary to illustrate the core planning decisions,
including stream selection, filtering, fusion, retrieval outputs, and the generation schema.

\begin{tcolorbox}[
  title={Example~1: Retrieval Planning Contract},
  label=ex:planning_contract,
  fontupper=\small,
  left=2mm,
  top=1mm,
  bottom=1mm,
  colback=gray!3
]
\begin{lstlisting}[
    language=json,
    basicstyle=\ttfamily\footnotesize
]
{
 "streams": [ // (1)
  "transcription", "speaker"
 ],
 "filters": { // (2)
  "text": "employment",
  "speaker": "SPEAKER_02"
 },
 "fusion": { // (3)
  "anchor": "transcript"
 },
 "output": { // (4)
  "return_fields": [
   "start","end","speaker","text"
  ]
 },
 "answer_schema": { // (5)
  "properties": {
   "answer": {
    "type": "string",
    "enum": ["A", "B", "C", "D"]
   }
  },
  "required": ["answer"],
 }
}
\end{lstlisting}
\end{tcolorbox}

\subsubsection{Stage 3: Structured Retrieval}
Given the retrieval plan $\Theta(q)$ from Stage~2, a rule-based SQL generator deterministically compiles it into an executable merged SQL query $Q(\Theta(q))$,
constructed using stream selection, filtering, fusion, and the output contract (items 1--4 in Stage~2).
The query is executed against the audio database $D(a)$ for the target recording $a$, yielding the retrieved segments $R(q,a)$:
\begin{equation}
R(q,a) = \mathrm{Exec}\big(Q(\Theta(q)),\, D(a)\big),
\end{equation}
where $\mathrm{Exec}(\cdot,\cdot)$ denotes a deterministic database execution operator.

Example~2 illustrates how the retrieval plan in Example~1 is compiled into a structured SQL query.
The \texttt{WITH} clause defines stream-specific Common Table Expressions (CTEs) ((A), (C)) with their corresponding filters ((B), (D)),
while the final \texttt{SELECT} projects the requested fields (E) and merges the streams via a temporal join (F),
realizing the fusion strategy specified in the plan.
We use a simple nearest-neighbor temporal fusion strategy based on
timestamp alignment; implementation details are provided in Appendix~\ref{appdx:temporal_fusion}.
Because each stream is translated into an independent CTE with its own filters, the resulting query is modular and naturally scalable to additional modalities.
We adopt a simple keyword-based retrieval mechanism in this work,
as our goal is to isolate the effect of explicit retrieval planning for complex long-form audio reasoning,
including both cross-modality and single-modality inference.

\begin{tcolorbox}[
  title={Example~2: Simplified Merged SQL},
  label=ex:sql_example,
  fontupper=\small,
  left=2mm,
  top=1mm,
  bottom=1mm,
  colback=gray!3
]
\begin{lstlisting}[
    language=SQL,
    basicstyle=\ttfamily\footnotesize
]
WITH
tx AS (  -- (A) transcript stream
  SELECT start, end, text
  FROM transcription
  WHERE text ILIKE '%employment%'
    -- (B) text filter
),
sp AS (  -- (C) speaker stream
  SELECT start, end, label
  FROM speaker
  WHERE label = 'SPEAKER_02'
    -- (D) speaker filter
)
SELECT   -- (E) output projection
  tx.start, tx.end, sp.label, tx.text
FROM tx
JOIN sp ON temporal_overlap(tx, sp);
  -- (F) stream fusion
\end{lstlisting}
\end{tcolorbox}

\subsubsection{Stage 4: Answer Generation}
The execution of the merged SQL query yields a set of relevant segments $R(q,a)$ from the audio database for the target recording $a$.
These segments are provided to a generation LLM together with an explicit output schema,
and the model produces the final answer while adhering to the schema constraints specified during planning.

% \subsection{Dump Audio information into Database}

\begin{figure}[t]
  \centering
  \includegraphics[width=\linewidth]{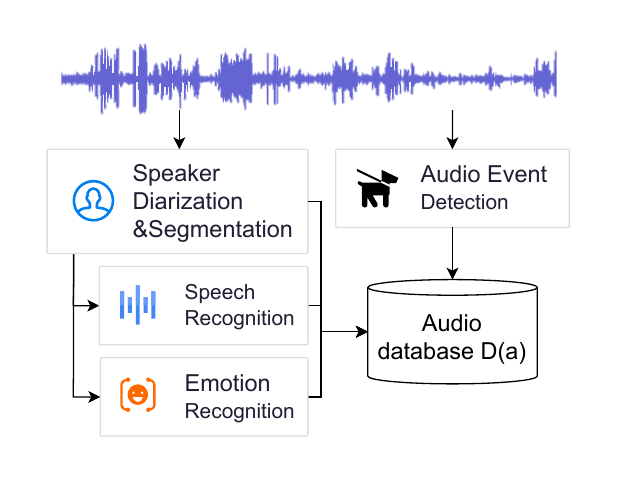}
  \caption{
Audio database construction.
Raw audio is processed by task-specific modules to construct a structured, time-aligned audio database $D(a)$ consisting of modality-specific metadata streams.
% Overview of the audio analysis pipeline. Raw audio is processed by multiple task-specific modules,
% and the resulting metadata are stored in a unified audio database.
% For emotion and speaker embedding extraction, segment boundaries are obtained using a Connectionist Temporal Classification (CTC)-based ASR model.
  }
  \label{fig:detailed_audio_ingestion}
\end{figure}

\subsection{Audio Database Construction}
\label{ssec:audio_database_construction}

Figure~\ref{fig:detailed_audio_ingestion} illustrates the construction of the structured audio database $D(a)$ from raw audio.
The pipeline begins with speaker diarization, which produces a sequence of speaker-homogeneous temporal segments.
These diarization timestamps define the fundamental alignment units used throughout the database construction process.

Given these shared segment boundaries, speech transcription and emotion recognition are applied to exactly the same temporal spans.
As a result, transcript, speaker, and emotion streams share identical start and end times, as illustrated by the record examples in Table~\ref{tab:data_example}.
This design enables cross-stream alignment, allowing text content to be directly matched with speaker identity and emotional cues using simple timestamp-based joins.

In contrast, sound event streams are generated at a different temporal resolution using a sliding-window audio tagging approach.
Unlike speech-centric streams, acoustic events do not necessarily depend on conversational structure or speaker boundaries.
Therefore, event predictions are produced independently of diarization segments and stored with their own start and end times.
As with emotion, event labels are stored as label--score pairs in \texttt{JSONB} fields, following a unified schema.

\begin{table}[!tp]\centering
\caption{Examples of records stored in the audio database $D(a)$, where each stream consists of time-aligned records;
label–score pairs for \texttt{emotion} and \texttt{sound\_event} are shown in simplified form.% for readability.
}
\label{tab:data_example}
\resizebox{\linewidth}{!}{ % use this if the table is too large
\begin{tabular}{lcccc}\toprule
Stream &Start (s) &End (s)  &Example \\\midrule
\texttt{transcript} &20.50 &22.10  &He talks about it \\
\texttt{speaker} &20.50 &22.10  & SPEAKER\_07 \\
\texttt{emotion} &20.50 &22.10  &Neutral (0.58), $\cdots$ \\
\texttt{sound\_event} &22.00 &27.00  &Speech (0.87),$\cdots$ \\
\bottomrule
\end{tabular}
}
\end{table}

\section{Experimental Setup}

Our experiment is designed to evaluate whether explicit retrieval
planning enables reliable and scalable understanding of long-form audio in a zero-shot setting.
% Rather than treating performance degradation under long inputs as a monolithic failure,
We decompose evaluation into two levels.
\emph{Base tasks} assess fundamental audio understanding capabilities, including
semantic understanding, speaker diarization, emotion recognition, and sound
event detection, while controlling for input duration.
\emph{Advanced tasks} require additional reasoning beyond direct retrieval,
such as counting, temporal ordering, and compositional constraints across
modalities.
Motivated by the limitations of existing long-form audio benchmarks discussed in Section~\ref{ssec:audio_evaluate}, we design our evaluation to emphasize reproducibility, domain relevance, and scalable reasoning over extended audio.

\subsection{Datasets}

% \paragraph{Task taxonomy}
% We organize our evaluation into two levels: \emph{base tasks} and \emph{advanced tasks}.
% Base tasks are designed to isolate and evaluate a single core capability for long-form audio understanding,
% such as semantic understanding (QA and summarization), speaker diarization,
% emotion recognition, or sound event detection.
% Advanced tasks require inference beyond direct retrieval and reporting.
% This includes reasoning-intensive queries within a single modality
% as well as compositional queries combining multiple capabilities.

\begin{table*}[!tp]\centering
\caption{Representative query examples for each task.
These queries illustrate the user-level intent of each task,
while the full prompts used for LLM inference are provided in Appendix~\ref{appdx:user_query}}
\label{tab:query_example}
\resizebox{\linewidth}{!}{ % use this if the table is too large
\begin{tabular}{ll}\toprule
Task &Example Query \\
\midrule
\multicolumn{2}{c}{Base Task}\\
\midrule
QA-1 &Who is Bela and why was no single arm is able to knock him down? \\
MCQA &What did the woman do to try to get the man's attention? A) ... \\
Summarization &Please provide an abstractive summary of the meeting segment between 0 and 600 seconds. \\
Diarization &Perform speaker diarization between 300 and 600 seconds. \\
Emotion &Analyze the audio between 325.41 and 332.23 seconds and respond with the emotion \\
SED &Detect occurrences of the following sound event label(s): Flamenco \\
\midrule
\multicolumn{2}{c}{Advanced Task}\\
\midrule
Event Ordering &Determine the order of first occurrence for: (1) Music (2) Bird flight, flapping wings (3) Change ringing \\
Speaker count &You should count the number of speakers starting from 300 sec to 600 sec. \\
Speaker-Constrained QA &You should work on the utterance from speaker 439. What does the woman say to the executioner? A)... \\
\bottomrule
\end{tabular}
}
\end{table*}

\paragraph{Base tasks}
We use only publicly available datasets to construct an evaluation set for each capability for the reproduction purpose.
For semantic understanding, we construct long-form recordings from LibriSpeech~\cite{librispeech} and evaluate question answering using LibriSQA~\cite{librisqa}, which is built on the \texttt{train-clean-360} subset and includes both abstractive QA (QA-1) and multiple-choice QA (MCQA).
For speaker diarization (SD), we generate long-form recordings by cropping or concatenating meeting audio from AMI~\cite{ami}.
For summarization, we use AMI recordings segmented into non-overlapping 10-minute windows and generate reference summaries by prompting ChatGPT with the corresponding reference transcriptions to produce 5--7 sentence abstractive summaries.
Emotion recognition (ER) is evaluated on the test-1 split of MSP-Podcast~\cite{msp_podcast},
which contains natural podcast recordings annotated with categorical emotion labels.
For sound event detection (SED), we construct extended recordings from VoxPopuli~\cite{voxpopuli} and insert AudioSet~\cite{audioset} clips as target events, following the needle-in-a-haystack paradigm used in Gemini 1.5 evaluation.

To control evaluation scale, we generate one question per 5 minutes of audio, resulting in 1000 questions per task, except for summarization where we use 100 questions.
For SD and summarization, queries are generated by partitioning each recording into non-overlapping temporal windows sampled sequentially from start to end, using 5-minute windows for SD and 10-minute windows for summarization.
For long-form evaluation, we construct audio inputs with durations of 10, 30, 60, 300, and 540 minutes.
All questions are instantiated from task-specific templates, with representative query examples provided in Appendix~\ref{appdx:user_query}.
Although long-form inputs are constructed through concatenation, the underlying datasets are derived from real-world recordings. AMI consists of approximately one-hour meeting recordings with natural multi-speaker interactions, while MSP-Podcast is based on podcast audio with diverse conversational styles. As a result, the constructed inputs preserve realistic conversational structure and acoustic variability.

\paragraph{Advanced Tasks}
For advanced tasks, we construct evaluation datasets by transforming base-task datasets to require additional inference over the same audio.
Specifically, for SD, we replace base-task queries with questions that require counting the number of distinct speakers appearing in a 10-minute recording.
Similarly, for SED, we reuse the original SED evaluation data (six events per 30-minute recording), randomly selecting three events to form event-ordering questions based on annotated onset times.
Due to this formulation, each recording yields a single counting question; we therefore construct this evaluation using 100 recordings, resulting in 100 ordering samples.

We design compositional tasks that combine multiple base capabilities within
a single query.
In speaker-constrained question answering, a target speaker is specified and the
system must answer using only utterances produced by that speaker.
We generate both answerable and unanswerable cases by leveraging speaker annotations
associated with existing QA pairs: for answerable cases, the original speaker is
specified, while for unanswerable cases, a different speaker appearing in the same
recording is specified, requiring the system to abstain.
We construct this task using 60-minute recordings, which contain 3-4 speakers and enable meaningful speaker-constrained reasoning.\footnote{Data and code will be released upon acceptance.}

\begin{figure*}[t]
  \centering
  \includegraphics[width=\linewidth]{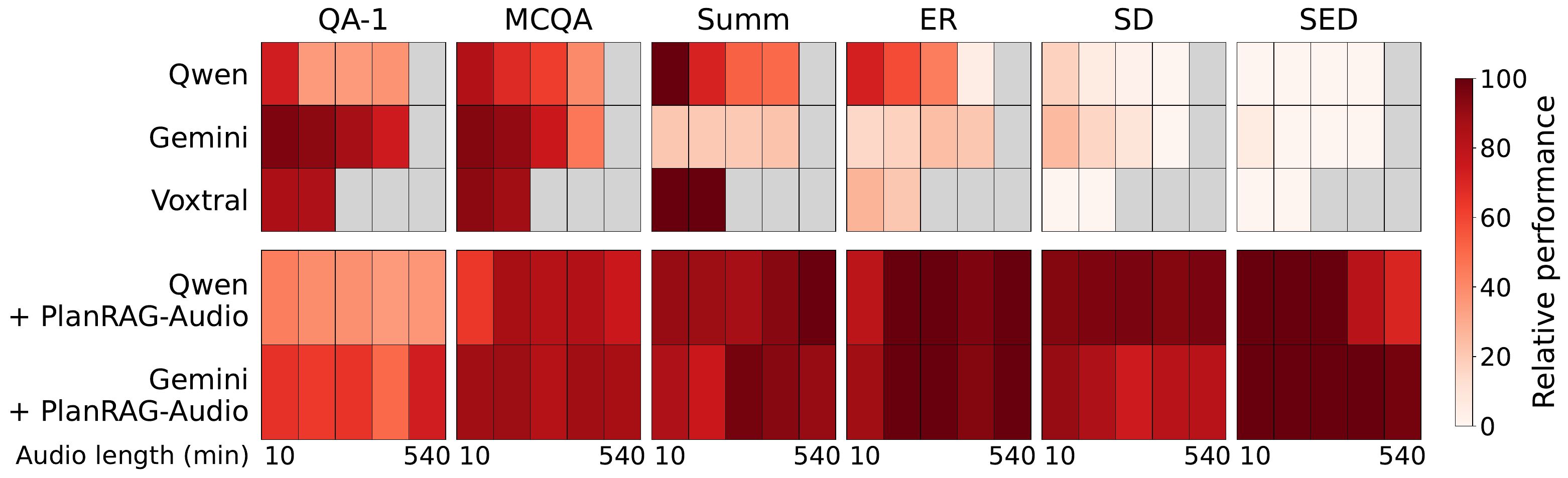}
  \caption{
Relative performance of base tasks under long-form audio inputs.
Results are for QA-1, MCQA, Summ, ER, SD, and SED across audio context lengths (10, 30, 60, 300, and 540 minutes).
All scores are normalized to a 0--100 scale per task, with error-based metrics inverted
(e.g., $100-\mathrm{DER}$) so that higher values indicate better performance.
Gray cells indicate unsupported or failed settings.
Exact values are provided in Appendix~\ref{appdx:detailed_results}.
}
% \caption{Performance of base speech tasks degrades as audio context length increases for vanilla LLMs,
% while retrieval-based abstraction (PlanRAG-Audio) stabilizes performance across tasks and time scales.}
  \label{fig:result_base_task}
\end{figure*}

\begin{table}[!tp]\centering
\caption{Model configurations used in our experiments}
\label{tab:model_table}
\resizebox{\linewidth}{!}{ % use this if the table is too large
\begin{tabular}{lrrr}\toprule
\textbf{Model} &\textbf{Version/Variation} &\textbf{Params} \\\midrule
Qwen &Qwen3-4B-Instruct-2507 &4B \\
Gemini &Gemini 2.5 Flash &undisclosed \\
Voxtral &Voxtral-Mini-3B-2507 &5B \\
ASR &OWSM-CTC v4 medium &1.01B \\
SED &BEATs iter3+, AS2M finetuned &90M \\
SD &Pyannote, community-1 & 8.1M \\
ER &Odyssey 2024 SER baseline  &316M \\
\bottomrule
\end{tabular}
}
\end{table}
\subsection{Models}
Table~\ref{tab:model_table} summarizes the models and configurations used in our experiments.
We adopt OWSM-CTC v4 as the ASR backbone due to its connectionist temporal classification (CTC)-only design, which avoids language-model memorization and enables fast, fully open multilingual recognition.
For audio understanding tasks, we employ standard pretrained models for SD, SED, and ER, while using \texttt{Qwen3-4B-Instruct} as the primary generation model.
In Figure~\ref{fig:result_base_task}, the Qwen row without PlanRAG-Audio serves as the baseline without planning, where the full structured audio database is provided directly to the LLM without retrieval planning or selective stream filtering.
As long-context baselines, we evaluate Gemini~2.5~Flash and Voxtral~\cite{voxtral} using their default inference settings, where the full audio input is provided without retrieval or segmentation.

\subsection{Metrics}

We report task-specific metrics following standard evaluation protocols:
Rouge-L for QA-1,
accuracy for MCQA,
Rouge-L for summarization,
Diarization Error Rate (DER) for SD,
macro F1 for ER,
and event-level F1-score~\cite{sed_metrics} with a fixed 5-second onset tolerance for SED.
% To enable unified visualization, all metrics are normalized to a 0–100 scale, with error-based metrics inverted $(100-\mathrm{DER})$.
% The topline for each task is defined as an oracle upper bound obtained by applying each pretrained model directly to the ground-truth–aligned segment.
% We report relative performance in the main figures, with absolute values in Appendix~\ref{appdx:detailed_results}.
% For advanced tasks, we report
% exact-match accuracy for speaker counting,
% Spearman's rank correlation for event ordering,
% and accuracy with abstention for speaker-constrained MCQA.
% Outputs that cannot be parsed are treated as incorrect.
% For MCQA, we report accuracy over parseable outputs; end-to-end results are in Appendix~\ref{appdx:detailed_results}.
% We report task-specific metrics following standard evaluation protocols.
% For QA-1 (LibriSQA Part~I), we evaluate generated answers using Rouge-L, while accuracy is reported for MCQA (Part~II).
% For summarization, we evaluate system outputs using Rouge-L against reference summaries generated by ChatGPT from the ground-truth transcript of the target time span specified in each query.
% For SD, we report Diarization Error Rate (DER).
% For ER, we report macro F1-score.
% For SED, we report event-level F1-score~\cite{sed_metrics} using a fixed 5-second onset tolerance without enforcing end-time localization.
To enable unified visualization across tasks, we normalize all metrics to a common 0--100 scale.
For metrics where higher is better, scores are linearly scaled by the task-specific topline; for error-based metrics such as DER, we apply $(100-\mathrm{DER})$ before normalization.

The topline for each task is defined as an oracle upper bound obtained by applying a pretrained model directly to the ground-truth–aligned segment without long-form context or retrieval.
Specifically, QA and summarization toplines are computed by applying Qwen3-4B model to the ground-truth answer text or transcript segment;
SD and ER toplines use the corresponding pretrained models.
For SED, BEATs performs clip-level classification without explicit timestamps, so the topline is computed on the evaluation window using macro F1-score.
We report relative performance in the main figures, with absolute values provided in the Appendix~\ref{appdx:detailed_results}.

% For advanced tasks, we report exact-match accuracy for speaker counting, Spearman's
% rank correlation for sound event ordering, and accuracy for
% speaker-constrained MCQA, with abstention accuracy reported for unanswerable cases.
% For all tasks, outputs that do not conform to the required response format or cannot be successfully parsed are treated as incorrect.

For advanced tasks, we report exact-match accuracy for speaker counting, Spearman's rank correlation for sound event ordering, and accuracy for speaker-constrained MCQA, with abstention accuracy for unanswerable cases.
For all tasks, outputs that do not conform to the required response format or cannot be parsed are treated as incorrect.
For QA, since it is sensitive to retrieval quality, we report accuracy over parseable outputs to isolate reasoning given retrieved evidence; end-to-end results are reported in the Appendix~\ref{appdx:detailed_results}.

\section{Results}

\begin{table}[t]
\centering
\caption{
Average number of input tokens passed to the LLM for MCQA with 60-min recordings.
}
\label{tab:token_length_simple}
\begin{tabular}{l r}
\toprule
Model & Avg. Tokens \\
\midrule
Gemini & 115.2k \\
Gemini + PlanRAG-Audio & 0.9k \\
Qwen + PlanRAG-Audio & 1.2k \\
\bottomrule
\end{tabular}
\end{table}

\subsection{Base Task Results}
% Compressed version
The evaluation results for base tasks are shown in Figure~\ref{fig:result_base_task}.
Without retrieval planning, performance degrades as audio duration increases for both Qwen and Gemini, with audio-only tasks such as SD, ER, and SED being particularly challenging.
This is evidenced by the Qwen row in Figure~\ref{fig:result_base_task}, where all database contents are passed to the LLM without planning, confirming that performance degradation stems from the absence of selective retrieval rather than model capacity alone.
Voxtral performs well on text-centric tasks but fails on non-text-based ones.
In contrast, PlanRAG-Audio stabilizes performance across input lengths by retrieving only query-relevant segments, effectively decoupling inference cost from raw audio duration and enabling reasoning over audio-derived signals.

Despite Gemini's long-context support, we observe notable degradation on long recordings, especially for speaker diarization.
Although the maximum output length is uniformly set to 4096 tokens, Gemini often produces incomplete or malformed outputs, causing parsing failures.
Across audio durations from 10 to 540 minutes, 17.92\% of diarization outputs cannot be successfully parsed and are therefore counted as incorrect.
As shown in Table~\ref{tab:token_length_simple}, Gemini processes the full speech input (e.g., 115k+ tokens for 60 minutes), whereas PlanRAG-Audio reduces the effective input to ~1k tokens via retrieval, keeping the LLM input size nearly constant and avoiding long-form accuracy degradation.

To better interpret these results, we decompose errors into three components: upstream perception, retrieval, and planning or generation. The topline results represent an upper bound determined by the pretrained perception models; the gap between topline and parseable reflects retrieval errors, while the gap between parseable and end-to-end captures planning and formatting failures. Detailed numerical results are provided in Appendix~\ref{sec:error_decomposition}.

\begin{table}[t]
\centering
\caption{
Performance on single-modality reasoning tasks.
Speaker counting and event ordering are evaluated using exact match accuracy and Spearman’s rank correlation, respectively.
}
\label{tab:single_modality_reasoning}
\resizebox{\linewidth}{!}{
\begin{tabular}{lcc}
\toprule
Model
& Speaker Count & Event Order \\
\midrule
Voxtral        & 9.17 & -0.10 \\
Gemini         & 14.20 & 0.30  \\
$\quad$+ PlanRAG-Audio      & 69.40 & 0.68  \\
Qwen & 35.16 & 0.11 \\
$\quad$+ PlanRAG-Audio  & 36.66 & 0.34  \\
\bottomrule
\end{tabular}
}
\end{table}
% Preamble:
% \usepackage{booktabs}
% \usepackage[table]{xcolor}

\begin{table}[t]
\centering
\caption{
% Speaker-constrained multiple-choice QA (\textbf{QA-2}).
% We report \textbf{QA accuracy (QA Acc.)} for answerable cases,
% \textbf{abstention accuracy (Abst. Acc.)} for non-answerable cases,
% and the average number of input tokens passed to the LLM.
% \textbf{SC} indicates whether speaker constraints are applied.
Speaker-constrained MCQA.
We report QA accuracy for answerable cases and abstention accuracy for non-answerable cases.
\textbf{SC} indicates whether speaker constraints are applied.
% Note: absent values are due to API rate limit.
% \markus{I would probably consider making part of the letters/words in the caption bold to better point the reader to the abbreviations.}
}
\label{tab:speaker_constrained_qa_mc}

\begin{tabular}{l c c c}
\toprule
Model & SC & QA Acc. & Abst. Acc.  \\
\midrule

\rowcolor{gray!30}
\multicolumn{4}{c}{\textit{Without PlanRAG-Audio}} \\

\rowcolor{gray!10}
Gemini &  & 58.83 & --  \\
\rowcolor{gray!10}
\phantom{Gemini} & \checkmark & 68.13 & 0.54  \\

\rowcolor{pink!70}
\multicolumn{4}{c}{\textit{With PlanRAG-Audio}} \\

% --- Gemini (2-row merged look) ---
\rowcolor{pink!25}
Gemini &  & 65.00 & -- \\
\rowcolor{pink!25} & \checkmark & 70.96 & 94.90 \\

\hdashline
% --- Qwen + PlanRAG (2-row merged look) ---
\rowcolor{pink!25}
Qwen  &  & 65.09 & --  \\
\rowcolor{pink!25}
\phantom{Qwen} & \checkmark & 67.59 & 82.20 \\
\bottomrule

\end{tabular}
% }
\end{table}

\subsection{Advanced Task Results}

\paragraph{Single-modality tasks}
Table~\ref{tab:single_modality_reasoning} reports results on reasoning-intensive single-modality tasks, speaker counting and sound event ordering.
Without PlanRAG-Audio, all models perform poorly on speaker counting,
while Gemini exhibits limited but non-trivial capability on event ordering, reflecting its long-context modeling capacity, and Voxtral fails to capture temporal order.
Applying PlanRAG-Audio substantially improves both tasks, most notably for Gemini,
where speaker counting accuracy increases from 14.20\% to 69.40\% and Spearman’s rank correlation for event ordering increases
from 0.30 to 0.68.
These gains stem from retrieval planning externalizing temporal structure and symbolic attributes:
speaker identifiers and event timestamps are explicitly provided in the retrieved results, reducing complex inference to counting unique labels or sorting by timestamp, tasks that are straightforward for the LLM but challenging when reasoning directly over raw audio.
% These gains stem from how retrieval planning restructures reasoning:
% for speaker counting, the retrieved results explicitly include speaker identifiers for each segment, reducing the task to counting the number of unique speaker labels,
% while for event ordering, the LLM is provided with event labels and their associated start and end timestamps and only needs to sort them by temporal order.
% Compared to long-context models that must implicitly infer speaker identities or event boundaries directly from raw audio,
% PlanRAG-Audio externalizes temporal structure and symbolic attributes through retrieval,
% thereby transforming the original reasoning task into a simpler and reliable post-retrieval problem for the LLM.

% Table~\ref{tab:single_modality_reasoning} reports results on single-modality reasoning-intensive tasks, including speaker counting and sound event ordering.
% Without PlanRAG-Audio, both Voxtral and Gemini perform poorly on these tasks, indicating that large-context audio models alone struggle to aggregate or temporally order information over long audio segments.
% In contrast, applying PlanRAG-Audio leads to substantial improvements, particularly for Gemini, where speaker count accuracy increases from 10\% to 80\% and event ordering performance improves significantly.
% These results demonstrate that explicit retrieval planning is crucial for enabling reliable reasoning over long-form audio, even in single-modality settings.

\paragraph{Compositional tasks}
Table~\ref{tab:speaker_constrained_qa_mc} reports results on speaker-constrained multiple-choice QA.
Without PlanRAG-Audio, Gemini shows limited abstention ability under speaker constraints.
With PlanRAG-Audio, QA accuracy is preserved while abstention accuracy is substantially improved, reaching 94.90\% for Gemini.
Qwen with PlanRAG-Audio achieves comparable constrained QA accuracy (67.59\%) and 82.20\% abstention accuracy.
These results demonstrate that PlanRAG-Audio correctly identifies the speaker specified in the question and retrieves evidence exclusively from the corresponding speaker stream, enabling reliable speaker-conditioned reasoning and abstention without task-specific SQL or handcrafted rules.
% Without PlanRAG-Audio, Gemini achieves reasonable QA accuracy but requires the full long-form context and fails to abstain in non-answerable cases.
% With PlanRAG-Audio, retrieval planning preserving QA accuracy and enabling abstention.
% For Qwen+PlanRAG-Audio, incorporating speaker constraints further improves reliability, achieving 81\% abstention accuracy.
% Importantly, these gains are obtained without introducing task-specific SQL queries or retrieval logic:
% the same retrieval planning framework generalizes across compositional tasks by reusing stream-level retrieval primitives.
% This demonstrates the robustness of PlanRAG-Audio in supporting diverse task compositions with minimal task-specific engineering.
% With PlanRAG-Audio, speaker constraints are externalized through retrieval planning by restricting evidence to utterances associated with the specified speaker,
% making the absence of supporting evidence explicit to the LLM.
% As a result, Qwen+PlanRAG-Audio achieves high abstention accuracy (81\%), while Gemini+PlanRAG-Audio still tends to over-generate answers, reaching only 24\% abstention accuracy.
% These results indicate that compositional failures primarily arise from missing speaker attributes rather than insufficient semantic understanding,
% and that retrieval planning enables this otherwise implicit constraint to be made explicit and verifiable.

\section{Conclusion}

We presented \textbf{PlanRAG-Audio}, a planning-based retrieval-augmented generation framework for scalable long-form audio understanding.
By planning which modalities, temporal spans, and output requirements are needed and retrieving only the relevant data.
%By planning and retrieving data for relevant modalities, temporal spans, and output requirements, PlanRAG-Audio enables reasoning without processing audio recordings.
% Across a wide range of base and advanced tasks, our experiments show that retrieval planning stabilizes performance as audio duration increases.
PlanRAG-Audio expresses complex cross-domain audio reasoning within a unified framework, without relying on task-specific prompts or bespoke query logic, and naturally extends to new modalities and longer audio streams.

\section*{Limitations}

Our evaluation of Gemini is constrained by practical limitations of the API, including unstable long-context handling and frequent formatting failures that prevent reliable evaluation.

We adopt a simple keyword-based retrieval mechanism to isolate the effect of retrieval planning. While more expressive retrievers may improve recall, our empirical results (Appendix~\ref{sec:semantic_search}) suggest that retrieval planning plays a more critical role than the choice of retriever in our setting.

Our framework depends on pretrained perception modules (e.g., ASR, SD, ER, and SED), and is therefore bounded by their accuracy. The primary goal of this work is not to optimize these components, but to demonstrate that long-form audio understanding can be effectively reformulated as a planning-based retrieval problem.

The preprocessing stage introduces additional computational cost, which is amortized across queries but may limit real-time applicability (see Appendix~\ref{sec:preprocessing_cost}).

From an ethical and risk perspective, PlanRAG-Audio does not introduce new modeling assumptions beyond those of the pretrained components it relies on.
As a result, potential risks present in pretrained models also apply to our framework.

\section*{The Use of AI Assistant}
Parts of this manuscript were edited for clarity and language using an AI-based writing assistant.
The authors take full responsibility for the content. 
\section*{Acknowledgments}

Experiments of this work used the Bridges2 system at PSC and the Delta/DeltaAI system at NCSA through allocations CIS210014 and IRI120008P from the Advanced Cyberinfrastructure Coordination Ecosystem: Services \& Support (ACCESS) program, supported by National Science Foundation grants \#2138259, \#2138286, \#2138307, \#2137603, and \#2138296. Thanks to the National Science and Technology Council (NSTC) for the funding under 114-2917-I-564-022.

% \clearpage
\bibliography{custom}

\clearpage
\appendix

\onecolumn
\section{User Queries}
\label{appdx:user_query}
\begin{tcolorbox}[
  title=Example: Query template for QA,
  fontupper=\footnotesize,
  left=2mm,
  width=\textwidth
]
\begin{lstlisting}
Given the context, answer the following question in a short sentence:
<question here>
\end{lstlisting}
\end{tcolorbox}

\begin{tcolorbox}[
  title=Example: Query template for Summary,
  fontupper=\footnotesize,
  left=2mm,
  width=\textwidth
]
\begin{lstlisting}
## Task
Please provide an abstractive summary of this meeting.

You should work on summarization starting from <start> sec to <end> sec.
Produce a concise, factual summary covering goals, key decisions, concerns,
and next steps.

Stay within 5-7 sentences.
## Answer
\end{lstlisting}
\end{tcolorbox}

\begin{tcolorbox}[
  title=Example: Query template for Speaker Diarization,
  fontupper=\footnotesize,
  left=2mm,
  width=\textwidth
]
\begin{lstlisting}
## Task
Perform speaker diarization for the provided audio segment spanning
<start> to <end> seconds in the original recording.

## Requirements
- Generate the following json 
```
[
  {"start": "start_time", "end": "end_time", "speaker": "speaker1"},
  {"start": "start_time", "end": "end_time", "speaker": "speaker2"},
    ...
]
```

## Answer
\end{lstlisting}
\end{tcolorbox}

\begin{tcolorbox}[
  title=Example: Query template for Emotion Recognition,
  fontupper=\footnotesize,
  left=2mm,
  width=\textwidth
]
\begin{lstlisting}
You are an emotion recognition model. Analyze the audio between <start> and <end>
seconds and respond with the emotion in the format
{"labels": ["Happy", "Angry"]}.
Return the most likely label(s).
\end{lstlisting}
\end{tcolorbox}

\begin{tcolorbox}[
  title=Example: Query template for Sound Event Detection,
  fontupper=\footnotesize,
  left=2mm,
  width=\textwidth
]
\begin{lstlisting}
You are a sound event localization (SED) model. Detect occurrences of the
following sound event label(s): <event label> in the audio clip.
Return JSON in the format
{"events": [{"label": "<event label>", "start": 0.0, "end": 0.0}]}.
Fewer events are preferred.
\end{lstlisting}
\end{tcolorbox}

\begin{tcolorbox}[
  title=Example: Query template for Speaker Count task,
  fontupper=\footnotesize,
  left=2mm,
  width=\textwidth
]
\begin{lstlisting}
## Task
You should count the number of speakers starting from <start> sec to <end>  sec.

## Requirements
- Generate the following json including the number of speakers in integer.

```
{"answer": <integer>}
```

## Answer
\end{lstlisting}
\end{tcolorbox}

\begin{tcolorbox}[
  title=Example: Query template for Event Ordering task,
  fontupper=\footnotesize,
  left=2mm,
  width=\textwidth
]
\begin{lstlisting}
You are given the following sound event labels:
(1) <label 1>
(2) <label 2>
(3) <label 3>
Determine the correct chronological order of these events based on their first
occurrence in the audio clip (<start> to <end> seconds). Return JSON in the
format {"order": [1, 2, 3]}.
\end{lstlisting}
\end{tcolorbox}

\begin{tcolorbox}[
  title=Example: Query template for Speaker Constrained MCQA task,
  fontupper=\footnotesize,
  left=2mm,
  width=\textwidth
]
\begin{lstlisting}
## Task
You should work on the utterance from speaker <speaker>.
If you cannot answer the question from the given speaker, just reply
"This question is not answerable."

## Question
Given the context, answer the following question in a short sentence:
<question here>

## Answer
\end{lstlisting}
\end{tcolorbox}

\section{Error Analysis}
\label{sec:error_decomposition}

\begin{table}[h]
\centering
\begin{tabular}{cccc}
\hline
Duration (min) & Topline & +PlanRAG (parseable) & +PlanRAG (e2e) \\
\hline
10  & 79.40 & 65.67 & 50.05 \\
30  & 77.94 & 67.23 & 51.63 \\
60  & 78.90 & 65.09 & 52.25 \\
300 & 77.06 & 63.87 & 50.95 \\
540 & 75.56 & 56.70 & 41.04 \\
\hline
\end{tabular}
\caption{Error decomposition across durations. The gap between topline and parseable reflects retrieval errors, while the gap between parseable and end-to-end reflects planning failures.}
\label{tab:error_decomposition}
\end{table}
\section{Detailed experimental results for base tasks}
\label{appdx:detailed_results}

\subsection{Top line results}

\begin{table}[!h]\centering
\caption{Topline results for abstractive QA (QA-1) and multiple-choice QA (MCQA)}
\begin{tabular}{lcccccc}\toprule
&\multicolumn{4}{c}{QA-1} &MCQA \\
\cmidrule(lr){2-5} \cmidrule(lr){6-6}
Duration(min) &BLEU &Rouge1 &Rouge2 &RougeL &acc \\\midrule
10 &29.03 &56.07 &38.67 &51.39 &79.4 \\
30 &30.39 &56.80 &39.50 &52.02 &77.94 \\
60 &31.27 &56.78 &40.35 &52.19 &78.9 \\
300 &31.15 &58.16 &40.6 &53.31 &77.06 \\
540 &30.54 &56.38 &39.4 &51.78 &75.56 \\
\bottomrule
\end{tabular}
\end{table}

\begin{table}[!h]\centering
\caption{Topline results for summarization}
\begin{tabular}{lccccc}\toprule
 & \multicolumn{4}{c}{Summ} \\
\cmidrule(lr){2-5}
Duration(min) &BLEU &Rouge1 &Rouge2 &RougeL \\\midrule
10 &9.21 &44.82 &14.74 &24.56 \\
30 &8.17 &42.66 &12.85 &23.08 \\
60 &7.82 &42.26 &12.61 &22.85 \\
300 &6.24 &36.99 &9.63 &20.04 \\
540 &4.61 &34.82 &7.56 &18.58 \\
\bottomrule
\end{tabular}
\end{table}

\begin{table}[!h]\centering
\caption{Topline results for speaker diarization, emotion recognition, and sound event detection}
\begin{tabular}{lcccccc}\toprule
&\multicolumn{2}{c}{SD} &\multicolumn{2}{c}{ER} &SED \\
\cmidrule(lr){2-3} \cmidrule(lr){4-5} \cmidrule(lr){6-6}
Duration(min) &DER &JER & Macro-f1 & Micro-f1 &acc \\\midrule
10 &10.55 &23.31 &26.48 &27.54 &50.89 \\
30 &12.64 &24.09 &19.32 &22.50 &48.48 \\
60 &12.42 &22.76 &14.56 &17.27 &49.03 \\
300 &12.42 & 22.51 &18.73 &19.26 &50.38 \\
540 &12.58 & 22.59 &20.13 &25.41 &51.46 \\
\bottomrule
\end{tabular}
\end{table}

\newpage
\subsection{Owsm + Qwen model}

\begin{table}[!h]\centering
\caption{Results for abstractive QA (QA-1) and multiple-choice QA (MCQA)}
\resizebox{\linewidth}{!}{
\begin{tabular}{lcccccccc}\toprule
& &\multicolumn{4}{c}{QA-1} &\multicolumn{2}{c}{MCQA} \\
\cmidrule(lr){3-6} \cmidrule(lr){7-8}
PlanRAG &Duration(min) &BLEU &Rouge1 &Rouge2 &RougeL &acc (parseable) & acc (end-to-end) \\\midrule
&10 &18.09 &42.41 &26.47 &37.66 &66.24 & 61.80 \\
&30 & 6.48 &20.18 & 8.26 &18.15 &53.13 & 31.34 \\
&60 & 6.57 &20.10 & 8.55 &18.28 &48.60 & 20.73 \\
&300 & 7.44 &22.15 & 9.77 &19.90 &30.69 & 8.33 \\
&540 & 6.28 &20.23 & 8.14 &18.15 &- & -\\ \midrule
$\checkmark$ &10 &18.35 &40.02 &25.40 &36.51 & 65.67 & 50.05 \\
$\checkmark$ &30 &17.05 &38.34 &23.64 &34.63 &67.23 & 51.63 \\
$\checkmark$ &60 &17.12 &38.29 &24.10 &34.64 &65.09 & 52.29 \\
$\checkmark$ &300 &17.09 &38.30 &23.76 &34.52 &63.87 & 50.95 \\
$\checkmark$ &540 &12.73 &30.24 &16.93 &26.91 &56.70 & 41.04 \\
\bottomrule
\end{tabular}
}
\end{table}

\begin{table}[!h]
\centering
\caption{Results for summarization (Summ), speaker diarization (SD), emotion recognition (ER), and sound event detection (SED) with OWSM $+$ Qwen model.}
\resizebox{\linewidth}{!}{
\begin{tabular}{lcccccccccccc}\toprule
& &\multicolumn{4}{c}{Summ} &\multicolumn{2}{c}{SD} &\multicolumn{2}{c}{ER} &\multicolumn{2}{c}{SED} \\
\cmidrule(lr){3-6}
\cmidrule(lr){7-8}
\cmidrule(lr){9-10}
\cmidrule(lr){11-12}
PlanRAG &Duration(min) &BLEU &Rouge1 &Rouge2 &RougeL &DER &JER & Macro-f1 & Micro-f1 &F1 (1 sec) &F1 (5sec) \\\midrule
&10 & 9.10 &43.52 &13.79 &24.48 &85.91 &82.08 &15.40 &19.13 &0 &0 \\
&30 & 5.22 &27.11 & 6.17 &16.37 &95.09 &94.01 &10.81 &12.55 &0 &0 \\
&60 & 2.33 &18.55 & 2.96 &11.83 &98.61 &97.68 & 7.29 & 8.25 &0 &0 \\
&300 & 1.37 &15.09 & 2.32 &10.07 &99.70 &99.58 & 0.91 & 0.95 &0 &0 \\
&540 & - &- & - &- &-     &-     & - & -&- &- \\ \midrule
$\checkmark$ &10 & 8.43 &40.02 &11.55 &22.13 &16.16 &25.24 &18.96 &21.83 &14.08 &73.94 \\
$\checkmark$ &30 & 6.33 &37.27 & 9.55 &20.46 &16.93 &23.66 &22.48 &24.85 &16.01 &55.27 \\
$\checkmark$ &60 & 5.97 &35.46 & 8.81 &19.82 &22.35 &27.72 &20.56 &23.51 &16.01 &49.76 \\
$\checkmark$ &300 & 5.18 &33.29 & 7.43 &18.72 &24.44 &28.49 &18.63 &22.65 &11.44 &41.04 \\
$\checkmark$ &540 & 4.25 &33.49 & 6.96 &18.42 &26.34 &30.03 &20.94 &25.55 &10.08 &35.94 \\
\bottomrule
\end{tabular}
}
\end{table}

\newpage
\subsection{Gemini results}

\begin{table}[!h]\centering
\caption{Gemini results for abstractive QA (QA-1) and multiple-choice QA (MCQA)}
%\resizebox{\textwidth}{!}{ % use this if the table is too large
\begin{tabular}{lcccccccc}\toprule
& &\multicolumn{4}{c}{QA-1} &\multicolumn{2}{c}{MCQA} \\
\cmidrule(lr){3-6}
\cmidrule(lr){7-8}
PlanRAG &Duration(min) &BLEU &Rouge1 &Rouge2 &RougeL &acc (parseable) & acc (end-to-end) \\\midrule
&10 &25.92 &52.71 &36.47 &48.59 &74.45 &74.30 \\
&30 &25.74 &52.11 &35.97 &48.09 &70.84&70.56 \\
&60 &23.76 &48.70 &33.66 &45.32 &59.00&58.83 \\
&300 &18.90 &43.10 &27.16 &39.62 &35.29&35.29 \\
&540 &- &- &- &- &- &-\\
\midrule
$\checkmark$ &10 &16.47 &36.63 &23.39 &34.12 &70.00&45.24 \\
$\checkmark$ &30 &15.65 &35.40 &22.14 &32.69 &69.53&47.21 \\
$\checkmark$ &60 &15.18 &36.07 &21.38 &33.70 &65.00&41.60 \\
$\checkmark$ &300 &11.48 &30.12 &16.09 &26.89 &67.62&42.40 \\
$\checkmark$ &540 &20.42 &41.24 &27.99 &37.61 &65.13&43.90 \\
\bottomrule
\end{tabular}
\end{table}

\begin{table}[!h]
\centering
\caption{Gemini results for summarization (Summ), speaker diarization (SD), emotion recognition (ER), and sound event detection (SED)}
\resizebox{\linewidth}{!}{ % use this if the table is too large
\begin{tabular}{lcccccccccccc}\toprule
& &\multicolumn{4}{c}{Summ} &\multicolumn{2}{c}{SD} &\multicolumn{2}{c}{ER} &\multicolumn{2}{c}{SED} \\
\cmidrule(lr){3-6}
\cmidrule(lr){7-8}\cmidrule(lr){9-10}\cmidrule(lr){11-12}
PlanRAG &Duration(min) &BLEU &Rouge1 &Rouge2 &RougeL &DER &JER & Macro-f1 & Micro-f1 &F1 (1 sec) &F1 (5sec) \\\midrule
&10 &0.00 &8.14 &2.55 &5.19 &79.47 &86.18 &3.27 &13.09 &3.00 &3.28 \\
&30 &0.00 &7.72 &1.79 &4.73 &86.43 &90.15 &3.25 &12.97 &0.23 &0.24 \\
&60 &0.00 &7.52 &1.37 &4.61 &92.68 &92.76 &4.01 &11.56 &0.14 &0.16 \\
&300 &0.00 &6.98 &1.24 &4.31 &100.00 &91.77 &3.85 &12.34 &0.02 &0.03 \\
&540 &- &- &- &- &- &- &- &- &- &- \\
$\checkmark$ &10 &6.27 &34.42 &9.37 &20.59 &19.69 &29.69 &20.89 &22.74 &21.77 &74.58 \\
$\checkmark$ &30 &4.10 &30.16 &6.87 &17.41 &26.23 &33.81 &27.82 &28.69 &19.07 &72.17 \\
$\checkmark$ &60 &7.84 &38.38 &11.06 &22.18 &35.35 &38.63 &19.62 &20.90 &13.53 &73.96 \\
$\checkmark$ &300 &4.53 &30.98 &6.56 &18.71 &29.23 &34.63 &19.83 &25.52 &7.86 &60.01 \\
$\checkmark$ &540 &3.57 &28.39 &5.65 &16.78 &29.54 &34.77 &25.88 &28.44 &13.40 &49.82 \\
\bottomrule
\end{tabular}
}
\end{table}

\newpage
\subsection{Voxtral results}

\begin{table}[!h]
\centering
\caption{Voxtral results for abstractive QA (QA-1) and multiple-choice QA (MCQA)}
\begin{tabular}{lccccccc}\toprule
&\multicolumn{4}{c}{QA-1} &\multicolumn{2}{c}{MCQA} \\\cmidrule(lr){2-5}\cmidrule(lr){6-7}
Duration(min) &BLEU &Rouge1 &Rouge2 &RougeL &acc (parseable) & acc (end-to-end)  \\\midrule
10 &22.34 &47.85 &31.59 &43.90 &73.17&73.10 \\
30 &22.32 &47.08 &30.79 &43.44 &68.84&68.36 \\
\bottomrule
\end{tabular}
\end{table}

\begin{table}[!h]\centering
\caption{Voxtral results for summarization (Summ), speaker diarization (SD), emotion recognition (ER), and sound event detection (SED)}
\resizebox{\linewidth}{!}{ % use this if the table is too large
\begin{tabular}{lcccccccccc}\toprule
&\multicolumn{4}{c}{Summ} &\multicolumn{2}{c}{SD} &\multicolumn{2}{c}{ER} & \multicolumn{2}{c}{SED} \\
\cmidrule(lr){2-5}
\cmidrule(lr){6-7}
\cmidrule(lr){8-9}
\cmidrule(lr){10-11}
Duration(min) &BLEU &Rouge1 &Rouge2 &RougeL &DER &JER & Macro-f1 & Micro-f1 &F1 (1 sec) &F1 (5 sec) \\\midrule
10 &12.14 &46.86 &16.53 &28.12 &100.00 &100.00 &5.70 &14.80 &0.00 & 0.00 \\
30 &9.52 &41.69 &13.42 &24.97 &100.00 &100.00 &3.97 &12.44 &0.00 & 0.00\\
\bottomrule
\end{tabular}
}
\end{table}

\section{Survey on Interspeech Paper}
\label{appdx:interspeech_survey}

% To ground our evaluation design in community practice, we analyzed papers from Interspeech main conferences between 2020 and 2025 based on their session titles and paper titles.
We analyze Interspeech 2020–2025 main conference topics to justify that our evaluation
tasks (speaker, emotion, event, and long-form speech understanding) reflect dominant
research directions in the speech community.
Table~\ref{tab:interspeech_papers} summarizes the resulting topic distribution, highlighting the prevalence of speech recognition, speaker-related tasks, and paralinguistic analysis.
This analysis informed our choice of evaluation domains in the main experiments.

\begin{table}[!h]\centering
\caption{Overview of research topics and their frequency in Interspeech 2020–2025 sessions.}
\label{tab:interspeech_papers}
% \resizebox{\linewidth}{!}{ % use this if the table is too large
\begin{tabular}{lrr}\toprule
\textbf{Topics} &\textbf{count} \\\midrule
Speech Recognition &3556 \\
Speaker and Language Identification &333 \\
Speech Recognition: Architecture, Search \& Linguistic Components &288 \\
Speech Perception, Production and Acquisition &273 \\
Spoken Language Processing: Translation, Retrieval and Resources &264 \\
Phonetics, Phonology and Prosody &244 \\
Spoken Dialog Systems and Conversational Analysis &225 \\
Paralinguistics and Affective Computing &148 \\
Speech, Voice and Hearing Disorders &116 \\ \midrule
Speech Coding and Enhancement &349 \\
Speech Synthesis and Spoken Language Generation &674 \\
Analysis of Speech and Audio Signals &363 \\
\bottomrule
\end{tabular}
% }
\end{table}
\newpage
\section{Audio Preprocessing Cost}
\label{sec:preprocessing_cost}

\begin{table*}[h!]
\centering
\begin{tabular}{cccccccc}
\hline
Duration (min) & RTF & Total (sec) & SD (sec) & ASR (sec) & ER (sec) & SED (sec) & Gemini e2e (sec) \\
\hline
10  & 0.06 & 37.25   & 6.66   & 12.88  & 3.58   & 3.64   & 3.49 \\
30  & 0.05 & 93.80   & 22.75  & 38.77  & 10.68  & 10.93  & 8.39 \\
60  & 0.05 & 184.18  & 52.27  & 76.75  & 21.21  & 21.70  & 9.76 \\
300 & 0.06 & 1036.34 & 406.06 & 384.55 & 105.43 & 119.68 & 49.71 \\
540 & 0.06 & 1986.00 & 866.33 & 686.95 & 185.64 & 221.69 & - \\
\hline
\end{tabular}
\caption{Preprocessing cost as a function of audio duration. The cost scales approximately linearly with input length.}
\label{tab:preprocessing_cost}
\end{table*}
\section{Temporal Fusion Details}
\label{appdx:temporal_fusion}

For a given base stream segment, we consider candidate segments from a target stream that temporally overlap with the base segment within a fixed tolerance window $\pm \tau$ seconds. Among these candidates, we select at most one segment whose temporal midpoint is closest to that of the base segment. Formally, for each base segment $b$ and target segment $t$, the midpoint distance is defined as:
\[
\left | \frac{\text{b.start} + \text{b.end}}{2} - \frac{\text{t.start} + \text{t.end}}{2} \right |.
\]
The target segment with the minimum midpoint distance is selected, and
the two segments are grouped into the same retrieval record. If no
target segment satisfies the temporal overlap constraint, the base
segment is retained without a matched target segment.

Unless otherwise specified, we use $\tau = 2.5$ seconds in all experiments.

\section{Semantic Search}
\label{sec:semantic_search}

\begin{table}[h!]
\centering
\begin{tabular}{ccc}
\hline
Duration (min) & Keyword Search & Vector Search \\
\hline
30  & 67.23 & 60.40 \\
540 & 56.07 & 57.39 \\
\hline
\end{tabular}
\caption{Comparison between keyword-based and vector-based retrieval. Results show that more expressive retrieval does not consistently improve performance, suggesting that retrieval planning plays a more critical role than the choice of retriever.}
\label{tab:retrieval_comparison}
\end{table}
\end{document}